# Seasonal Influenza Vaccination Hesitancy and Digital Literacy: Evidence from the European countries*

Martina Celidoni, Nita Handastya, Guglielmo Weber, Nancy Zambon

*Department of Economics and Management,*

*University of Padova - Via Del Santo 33, 35123 Padua, Italy*

## Abstract

This study documents the relationship between computer skills/digital literacy and influenza vaccination take-up among older adults in Europe during and after the COVID-19 pandemic. Using data from the Survey of Health, Aging and Retirement in Europe, we find a positive partial association between influenza vaccination take-up and two indicators of computer skills/digital literacy, *self-assessed pre-pandemic computer skills* and *having used a computer at work in any pre-pandemic job*. We do not estimate significant behavioural changes for individuals with better computer skills that may have been driven by spillover effects from the pandemic experience.

* Acknowledgments: This paper was developed within the project funded by Next Generation EU - "Age-It - Ageing well in an ageing society" project (PE0000015), National Recovery and Resilience Plan (NRRP) - PE8 - Mission 4, C2, Intervention 1.3". The views and opinions expressed are only those of the authors and do not necessarily reflect those of the European Union or the European Commission. Neither the European Union nor the European Commission can be held responsible for them.

This paper uses data from SHARE SCS 2 and wave 9 release 9.0.0 (https://www.doi.org/10.6103/SHARE.w9ca.900, (https://www.doi.org/10.6103/SHARE.w9.900). Pre-pandemic data refers to the SHARE wave 8 release 9.0.0 (https://www.doi.org/10.6103/SHARE.w8.900) and when necessary, is complimented with data from SHARE waves 1, 2, 3, 4, 5, 6, and 7 (https://www.doi.org/10.6103/SHARE.w1.900, https://www.doi.org/10.6103/SHARE.w2.900, https://www.doi.org/10.6103/SHARE.w3.900, https://www.doi.org/10.6103/SHARE.w4.900, https://www.doi.org/10.6103/SHARE.w5.900, https://www.doi.org/10.6103/SHARE.w6.900, https://www.doi.org/10.6103/SHARE.w7.900). Further information regarding methodological details can be found on Börsch-Supan et al. (2013).

## 1 Introduction

Respiratory viral infections pose a significant global health challenge, with influenza alone estimated to cause one billion cases annually, of which 3-5 million severe cases and between 290,000 and 650,000 influenza-related respiratory deaths (World Health Organization, 2023; Iuliano et al., 2018). Many European countries have adopted organized vaccination programs as a key preventive strategy against seasonal influenza epidemics. These vaccines are generally offered free of charge and are easily accessible through family doctors or healthcare authorities responsible for preventive services.

Typically, seasonal influenza vaccination campaigns in Europe run annually between autumn through winter. Although the target group varies across countries and/or regions, there is a consensus regarding the importance of these campaigns (Blank et al., 2018). Vaccination against seasonal (epidemic) influenza is particularly beneficial for the vulnerable populations, including individuals aged 65 or older, those with fragile health conditions, and groups at increased risk of exposure to or transmission of influenza virus, such as health care/essential workers (World Health Organization, 2022a). While strongly recommended, especially for older individuals, the decision to get vaccinated remains a personal choice influenced by behavioural and social factors, which can lead to low or stagnating uptake rates (World Health Organization, 2022b).

The concept of 'vaccine hesitancy' has emerged to describe the delay in acceptance or refusal to get vaccinated, despite its availability (MacDonald & the SAGE Working Group on Vaccine Hesitancy, 2015). This phenomenon is now considered one of the top ten global health threat by the World Health Organisation (World Health Organization, 2019). Research on vaccine hesitancy, particularly in the context of the COVID-19 pandemic has identified several key factors contributing to vaccine refusal. These include concerns about vaccine safety and its side effects (Fieselmann et al., 2022), low perceived benefit and doubts regarding its efficacy (Hofstra and Larson, 2023), mistrust in health authorities and the pharmaceutical industry, and lack of information or exposure to misinformation (Rizzo et al., 2017; Garett and Young, 2021).

Socio-economic barriers also play a role in vaccine hesitancy, especially among the poor elderly population. Additionally, also demographic characteristics, self-perceived health, and employment status are significantly correlated to vaccination uptake among poor elderly individuals (Veronese et al., 2023).

The COVID-19 pandemic has posed new challenges related to vaccine uptake. During the global lockdowns, most people were physically isolated and exposed to conflicting information about COVID specific vaccines by media and social networks. Related to this, Principe and Weber (2023) found that health information seeking behaviour played a crucial role in COVID-19 vaccination hesitancy among older individuals, highlighting the importance of digital health literacy.

In light of these findings, we use data gathered through the Survey of Health, Ageing and Retirement in Europe (SHARE), during and after the COVID-19 pandemic, to provide new evidence on the correlation between computer skills/digital literacy and the seasonal influenza uptake decision and investigate whether there have been behavioural changes that may have been driven by spillover effects from the pandemic experience.

The paper is organized as follows. In Section 2 we review the literature on vaccine uptake and hesitancy, in Section 3 we describe the date used, we then present the empirical strategy in Section 4, Section 5 comments our results; Section 6 concludes.

## 2. Literature

Vaccine hesitancy (VH) is a complex, time-varying and context specific phenomenon. It is influenced by factors such as complacency, convenience and confidence (MacDonald & the SAGE Working Group on Vaccine Hesitancy, 2015). In an attempt to formalise the consensus regarding determinants of VH, The World Health Organisation (2022) outlines domains of behavioural and social drivers of vaccination, proposing interventions aimed at increasing vaccine literacy.

The literature on vaccine hesitancy has explored a wide range of possible determinants, including socio-economic disparities (Veronese et al., 2023), perceived risk (Deiana et al., 2022), distrust in vaccines and in governments (Jennings et al., 2021), historical and cultural reasons (Binzel and Link, 2023), political polarisation (Dolman et al., 2023), and misinformation (DeStefano and Thompson, 2004; Davidson, 2017).

Evidences, more specifically on the COVID-19 pandemic, show that socio-economic disparities are determinants of vaccine uptake. For instance, education and income, marital status, age, and gender are associated with an individual attitude toward vaccination (Malik et al., 2020; Jain et al., 2021; Yasmin et al., 2021; Lee and Huang, 2022; Limbu et al., 2022; Richter et al., 2022; Sharma et al., 2023; Zintel et al., 2023). Other indicators such as belonging to a certain racial or ethnic group, poverty, level of subjective health, access to healthcare, and feeling of security in one's neighbourhood have been found to correlate with a higher level of vaccine refusal (Mustafa et al., 2022; Richter et al., 2022; Moon et al., 2023; Veronese et al., 2023).

A literature review by Siram et al. (2024) explored vaccine hesitancy during the COVID-19 pandemic through the behavioural economics approach. This approach allows for an interaction between psychological and economic factors. They name six types of bias that arose from the pandemic, namely: availability bias (of the newly available information regarding vaccinations), optimism bias (of their own physical health), illusion of control bias (over immediate factors believed to spare them), omission bias (belief that side effect is more severe than contracting the virus), confirmation bias (of one's preconceived belief), and negative bias in framing (negative news is easily accepted compared to positive news). The biases fit closely with evidences found on empirical studies. Among the psychological determinants explored include perceived risk in the vaccine (McSpadden, 2021; Sherman et al., 2021; Wolff, 2021; Deiana et al, 2022), distrust in vaccines and in government (Jennings et al., 2021: McSpadden, 2021), historical and cultural reason (Binzel and Link, 2023), political polarisation (Dolman et al., 2023), and misinformation (DeStefano and Thompson, 2004; Davidson, 2017). Sometimes, the determinants interact with one another and fuelled the anti-vaccination sentiment (Nuwarda et al., 2022).

We are interested in older individuals, who are a vulnerable group of the population in this context, not only for health-related reasons, but also because a larger fraction of older cohorts is not familiar with new technologies that might have conveyed misinformation and generated hesitancy in vaccination.

Indeed, the COVID-19 pandemic triggered a significant shift in communication methods, with daily activities such as news consumption, health consultations, and social interactions migrating to digital platforms due to global lockdown policies. However, this transition was not experienced equally across all population segments, bringing the so-called digital divide to the attention (Van Jaarsveld, 2020).

The digital divide exacerbated what has been termed the 'triple exclusion' faced by the elderly. Thus comprises: 1) a higher risk of mortality from COVID-19, 2) challenges in accessing high-quality information or services online, and 3) an increased likelihood of experiencing social isolation and loneliness (Xie et al., 2021; Zapletal et al., 2023). While the internet became a crucial remedy to these exclusions, it also presented new challenges in keeping the elderly safe. As COVID-19 vaccination campaigns began in 2021, digital literacy and computer skills became increasingly important. Official health communications competed with other sources such as news outlets and social media, often presenting contradictory information. The resulting 'infodemic' – an overabundance of information both online and offline – amplified social anxieties and trust issues, further exacerbating vaccine scepticism (Pertwee et al., 2022).

Digital skills among the elderly are important as they help them to assess the validity of the information received online and make informed decisions regarding important preventive medicinal practices such as vaccination. Digital skills thus become an essential tool for increasing vaccine literacy, which has been found to be a strong predictor of an individual's intention to vaccinate in the future (Isonne, 2024a, 2024b). The importance of digital literacy has been discussed by literature when investigating the role of online vaccine information seeking behaviour and vaccination uptake or intention to vaccinate (Zheng et al., 2022; Principe and Weber, 2023; Paimre et al., 2024; Zhang et al., 2024)

Digital literacy and vaccination against COVID-19 behaviour have studied using different empirical approaches. Zheng et al. (2022) used the Structural Equation Modelling based on the SOR (Stimulus-Organism-Response) framework using cross-sectional data from China and the US. Specifically, they aim to study the indirect relationship between online vaccine information seeking and vaccination intention during COVID-19 pandemic via perceived information overload, vaccine risk perception, and negative affective response. Paimre et al. (2024) used cross-sectional data from SHARE data focusing only on respondents in Estonia and aimed to study how technology, health information seeking and socioeconomic factors associated with the COVID-19 vaccination. Zhang et al. (2024) focused on the impact of internet health information seeking and COVID-19 behaviour by using the Chinese General Social Survey (CGSS) data with the use of OLS (Ordinary Least Square) and PSM (Propensity Score Matching) model. Principe and Weber (2023) used the SHARE data to study the elderly population in Europe and how health information seeking affected vaccination hesitancy. The study used an instrumental variable (IV) approach by using computer skills and past experience of using computer at work as instruments for online health information seeking.

We add to the literature on vaccine hesitancy and uptake decision by providing new evidence on the correlation between computer skills/digital literacy and the seasonal influenza vaccination take-up, to understand whether there have been spillover effects from the pandemic experience on the seasonal influenza vaccination decision.

## 3. Data

This analysis exploits data from SHARE, an ongoing longitudinal, multi-disciplinary, and cross-national European study. The survey contains current and retrospective information on health, socio-economic status, and social and family networks of individuals aged fifty and older in twenty European countries and Israel (Börsch-Supan et al., 2013).

The collection of SHARE data started in 2004, currently data from the ninth wave are available to the scientific community referring to the period 2021-2022. During the COVID-19 outbreak in March 2020, the regular data collection for the eighth wave was suspended. Shortly after, two additional surveys were conducted between June-August 2020 (SHARE Corona Survey 1, SCS1) and June-August 2021 (SHARE Corona Survey 2, SCS2) with the aim to collect data on health and socio-economic impacts of COVID-19 among SHARE respondents.

In our analysis, we will focus on two main waves, namely: SHARE SCS2, and wave 9 (2021-2022), since our aim is to document behavioural changes in seasonal flu vaccination uptake during and after the pandemic, especially related to pre-determined digital skills.

We select individuals aged 50+ from 27 countries who were interviewed in both SCS2 and Wave 9; our sample is composed by about 26,000 respondents.

Table 1. Descriptive statistics

| Variable | All N = 51,372 | | SCS2 N = 25,686 | | Wave 9 N = 25,686 | |
|---|---|---|---|---|---|---|
| | Mean | SD | Mean | SD | Mean | SD |
| Flu vaccination status: Vaccinated | 0.4153 | | 0.3808 | | 0.4499 | |
| Pre-pandemic computer skill: Good | 0.6101 | | | | | |
| Any per-pandemic computer job: Yes | 0.4634 | | | | | |
| Was eligible for flu vaccination in wave8: Yes | 0.8072 | | | | | |
| | | | | | | |
| Eligible for flu vaccination | | | 0.8627 | | 0.8663 | |
| | | | | | | |
| Health variables | | | | | | |
| Numbers of chronic diseases | 1.4701 | 1.1207 | 1.3751 | 1.1194 | 1.5649 | 1.1138 |
| Subjective health: Poor | 0.0704 | | 0.0459 | | 0.0949 | |
| Is trustful toward others: Yes | 6.4098 | 2.2351 | | | | |
| | | | | | | |
| Socio-economic variables | | | | | | |
| Is a female: Yes | 0.5669 | | | | | |
| Age at the time of interview | 71.3106 | 8.8796 | 71.2214 | 8.8757 | 71.3999 | 8.8826 |
| Highest education attained (ISCED) | | | | | | |
| None | 0.0188 | | | | | |
| Primary education | 0.2588 | | | | | |
| Secondary education | 0.4690 | | | | | |
| Tertiary education | 0.2534 | | | | | |
| Area of household building: Urban | 0.7121 | | | | | |
| Household size | 1.9915 | 0.9121 | 2.00728 | 0.9244 | 1.9758 | 0.8992 |
| Household's income quartile in wave 8 | | | | | | |
| First quartile | 0.2265 | | | | | |
| Second quartile | 0.2441 | | | | | |
| Third quartile | 0.2660 | | | | | |
| Fourth quartile | 0.2635 | | | | | |

| | | | |
|---|---|---|---|
| Never did any paid work before the pandemic: Yes | 0.3820 | | |

In Table 1 we report the descriptive statistics. Information regarding individual computer skills, past jobs that required the use of a computer, and eligibility to receive flu vaccination refers to wave 8 and are therefore pre-determined.

The sample shows that there is a rather high number of individuals who rated their computer skill to be good (61%), even though only a little bit less than half ever had any computer-using job (46%). The two binary indicators, ‘Pre-pandemic computer skill: Good’ and ‘Any per-pandemic computer job: Yes’, are constructed using pre-pandemic information about computer skills and computer use at work. The first indicator is based on a variable worded ‘How would you rate your computer skills? Would you say they are…’ with 6 possible responses, ‘Excellent’, ‘Very good’, ‘Good’, ‘Fair’, ‘Poor’, ‘I never used a computer’ (Wave 5 to wave 8). The responses are grouped into a dummy variable, with the first three marked as ‘Yes’ and the latter three as ‘No’. The second indicator, instead, is constructed using retrospective information from variables worded as ‘Does your current job require using a computer?’ (Wave 1 to wave 8, except wave 3) and ‘Have you ever used a computer at work?’ (Wave 7).

Eligibility for seasonal influenza vaccination has been defined according to the information reported in Table 2 about the EU/EEA-wide recommendation for the 2021-22 influenza season and captures institutional policy interventions aimed to protect the most vulnerable segments of the population. Table 1 shows that most individuals (81%) were already eligible to receive flu vaccination in wave 8.

Our outcome of interest is the individual vaccination uptake. SHARE gathers information on flu vaccination since wave 8 through the following question: ‘In the last year, did you have a flu vaccination?’. From Table 1, we can observe how influenza-vaccination uptake has increased after the pandemic, moving from 38% in SCS2 to 45% in Wave 9.

Table 3 shows also the differences in influenza-vaccination uptake between individuals having better/worse computer skills and those having any pre-pandemic computer job or not in the two periods analysed. We can see that vaccination uptake is significantly higher among those having computer skills or having had any per-pandemic computer job in all time periods. We can also observe that the increase in uptake in wave 9, with respect to SCS2, for those reporting having good pre-pandemic computer skills is higher than those reporting worse pre-pandemic computer skills.

Additional individual-level variables considered in our analysis and summarized in Table 1 are gender, age, education, and a binary indicator for ever done paid work. We further exploit household characteristics (income quartile, size, geographical area), and health related information (subjective health, chronic diseases) as set of control covariates. The never worked for pay indicator was generated from retrospective information worded ‘Have you ever done paid work?’ collected from wave 1 to wave 8. The response ‘Yes’ indicates that the individual had never worked for any payment, while ‘No’ means otherwise.

Table 1 shows that 57% of the respondents are females, with an average age at the time of the interview of 71.2 in SCS2 and 71.4 in Wave9, a medium-high educational level (72% of the respondents have a secondary or tertiary education), and who often worked for pay before the pandemic (only 38% Never did any paid work before the pandemic). Table 1 further shows that respondents are part of small households, with an average household size that decreases from 2.01 in SCS2 to 1.98 in Wave 9, and often resides in an urban area (71%).

Table 2. Seasonal influenza vaccination policy across EU countries

| Country | ≥ 18 years | ≥ 50 years | ≥ 55 years | ≥ 59 years | ≥ 60 years | ≥ 65 years | Funding/administration of the vaccine | Recommended vaccine product |
|---|---|---|---|---|---|---|---|---|
| Austria | R | | | | | | F/F | IIV4, aIIV4, QIV-HD |
| Belgium | | R | | | | R | F/F | IIV3, IIV4 |
| Bulgaria | R | | | | | | F/F | IIV4 |
| Croatia | | | | | | R | F/F | IIV4 |
| Cyprus | | | | | | R | F/F | aIIV4 |
| Czechia | | R | | | | R | F/F | IIV4 |
| Denmark | | | | | | R | F/F | IIV4, QIV-HD |
| Estonia | R | | | | | | F/F | IIV4 |
| Finland | | | | | | R | F/F | IIV4 |
| France | | | | | | R | F/F | IIV4, QIV-HD |
| Germany | | | | | R | | F/F | IIV4, QIV-HD |
| Greece | | | | | R | | F/F | IIV4 |
| Hungary | | | | | R | | F/F | IIV3 |
| Israel | | | | | | R | F/F | |
| Italy | | | | | R | | F/F | IIV4, aIIV4 |
| Latvia | | | | | | R | F/F | IIV4, aIIV4, cIIV4, rIIV4, QV-HD |
| Lithuania | | | | | | R | F/F | IIV4 |
| Luxembourg | | | | | | R | F/F | IIV4 |
| Malta | | | R | | | | F/F | IIV4 |
| The Netherlands | | | | | R | | F/F | IIV4 |
| Poland | | | R | | | | F/F | IIV4 |
| Romania | | | | | | R | F/F | IIV4 |
| Slovakia | | | | R | | | F/F | IIV4 |
| Slovenia | | | | | | R | F/F | IIV4 |
| Spain | | | | | | R | F/F | IIV4, aIIV3, aIIV4, cIIV4, QIV-HD |
| Sweden | | | | | | R | F/F | IIV4, QIV-HD |
| Switzerland | | | | | | R | F/F | IIV4, QIV-HD |

R: Recommended. Recommended vaccination is defined as the existence of a written recommendation in an official policy document stating that a particular population should receive seasonal influenza vaccine.
F: Funded.
F/F: Funding of the vaccine/Funding of the administration of the vaccine.
IIV3: Trivalent inactivated influenza vaccine;
IIV4: Quadrivalent inactivated influenza vaccine;
aIIV3: adjuvanted trivalent influenza vaccine;
aIIV4: adjuvanted quadrivalent influenza vaccine;
cIIV4: cell-derived inactivated quadrivalent influenza vaccine;
QIV-HD: high-dose quadrivalent influenza vaccine.
rIV4: recombinant quadrivalent influenza vaccine.

**Austria**: Recommended in all individuals, fully funded in those aged 6 months-15 years, in individuals living in retirement homes and long-term care facilities, in those aged >=60 years. Some healthcare insurers and employers offer (partly) funded influenza vaccination programmes.
**Belgium**: The vaccination is recommended in individuals aged >=65 years which is the priority group. Depending on vaccine availability, the recommendation also applied to healthy individuals aged >=50 years. Depending on the age group and risk group, funding may be partial or total.
**Bulgaria**: National programme for improvement of seasonal flu vaccine prophylaxis, 2019-2022 (in Bulgarian): https://www.strategy.bg/StrategicDocuments/View.aspx?lang=bg-BG&Id=1275 Free of charge in those aged 65 years and above until the 2021-22 influenza season.
**Czechia**: The Czech Vaccine Society recommends the vaccine in all adults over 50 years while the Czech National Immunisation Technical Advisory Group recommends the vaccine in all adults over 65 years.
**Denmark**: QIV-HD recommended in adults aged >=82 years.

**Estonia**: funded in those aged >=65 years. https://ta.vaktsineeri.ee/et/haigused-ja-vaktsiinid/vaktsineerimine-eestis/riiklik-immuniseerimiskava, https://ta.vaktsineeri.ee/et/taiskasvanutele/mille-vastu-saab-vaktsineerida; https://www.terviseamet.ee/sites/default/files/contenteditor/vanaveeb/Nakkushaigused/immunoprof/Lisa_2_vaktsiinid_ja_sihtruehmad.pdf. All adults (especially those aged >= 65 years) should be vaccinated against seasonal influenza: https://www.terviseamet.ee/sites/default/files/content-editor/vanaveeb/Nakkushaigused/immunoprof/Lisa_2_vaktsiinid_ja_sihtruehmad.pdf g
Finland: https://thl.fi/en/web/infectious-diseases-and-vaccinations/information-about-vaccinations/vaccination-programme-for-children-and-adults; Vaccines available during the 2021-22 season: https://thl.fi/en/web/infectious-diseases-and-vaccinations/vaccines-a-to-z/influenza-vaccine#where
**France**: QIV-HD only in those aged >=65 years
**Germany**: https://www.rki.de/DE/Content/Infekt/EpidBull/Archiv/2022/Ausgaben/04_22.pdf?__blob=publicationFile; QIV-HD in those aged 60 years and over.
**Italy**: aIIV4 and QIV-HD in those aged >=65 years.
**Israel: Vaccine type is not specified**
**Latvia**: https://www.spkc.gov.lv/lv/media/5827/download
**Lithuania**: Vaccine and its administration partially funded.
**Malta**: https://deputyprimeminister.gov.mt/en/phc/pchyhi/Pages/Vaccines.aspx
**Poland**: source: https://szczepienia.pzh.gov.pl/kalendarz-szczepien-2022-2/; https://szczepienia.pzh.gov.pl/bezplatne-szczepienia-przeciw-grypie-dla-doroslych/; Funding and vaccines: https://szczepienia.pzh.gov.pl/faq/komu-przysluguje-bezplatna-lub-refundowana-szczepionka-przeciw-grypie/, https://szczepienia.pzh.gov.pl/bezplatne-szczepienia-przeciw-grypie-dla-doroslych/ . Vaccine and its administration are partially funded (50%).
**Slovenia**: vaccination against influenza is recommended for everyone >6 months, but especially recommended and funded for the specific groups mentioned (children aged 6-23 months, pregnant women, chronic patients, older individuals aged 65 years and above).
**Sweden**: High-dose vaccine quadrivalent recommended for residents in long-term care facilities.
**Switzerland**: https://www.bag.admin.ch/dam/bag/en/dokumente/mt/infektionskrankheiten/grippe/empfehlung-grippeimpfung-kurz.pdf.download.pdf/empfehlungen-grippeimpfung-kurz-en.pdf

*Source: European Centre for Disease Prevention and Control (2023)*

Table 3. Seasonal flu vaccination uptake by computer skills/digital literacy

| | Pre-pandemic computer skills | | |
|---|---|---|---|
| | Good | Less than good | P-values |
| Take-up - SCS2 | 0.4062 | 0.3408 | 0.000 |
| Take-up - Wave 9 | 0.4761 | 0.4087 | 0.000 |
| | Any pre-pandemic computer job | | |
| | Yes | No | P value |
| Take-up - SCS2 | 0.4249 | 0.3425 | 0.000 |
| Take-up - Wave 9 | 0.4921 | 0.4134 | 0.000 |

As regard health variables, we can see in Table 1 that respondents report a low number of chronic conditions and only rarely assess their health as poor. Indeed, the average number of chronic diseases is, in SCS2, about 1.4 out of 7 chronic diseases listed in the survey (hip fracture, diabetes or high blood sugar, high blood pressure or hypertension, heart attack or other heart problem, chronic lung disease, cancer or malignant tumor, other illness or health condition). This number slightly increases to 1.6 after the pandemic. Subjective health, instead, is based on a variable worded 'Would you say your health is…' with 5 possible responses, 'Excellent', 'Very good', 'Good', 'Fair', 'Poor'. The percentage of respondents who reported poor health more than doubled between the waves shifting from 4.6% to 9.5%. We also included a variable to capture an individual's trust in others. The information is collected only once when the respondent enters the sample for the first time. The question is phrased as follows '*Generally speaking, would you say that most people can be trusted or that you can't be too careful in dealing with people*?' with an answer scaled from 0 to 10, where 0 means 'you can't be too careful' and 10 means that 'most people can be trusted'. On average, the response is about 6.5 out of 10.

## 4. Empirical strategy

To investigate the association between computer skills/digital literacy and vaccine hesitancy, we consider the following model:

$$y_i = \alpha + \beta_1 \text{ Pre-pandemic computer skills}_i + \beta_2 \text{ Any pre-pandemic computer job}_i + X'_i\gamma + e_i. \qquad (1)$$

where $y_{it}$ is the outcome of interest – seasonal flu vaccination uptake - for individual *i* and *Pre-pandemic computer skills*$_i$ and *Any pre-pandemic computer job*$_i$ are our binary indicators, capturing computer skills/digital literacy. The matrix $X'_{it}$ contains the covariates we include in our model: characteristics such as age, gender, country of residence, educational level, household's income quartile dummies, household's size, an indicator for individuals who have never worked for pay, whether the individual lives in a urban or rural area, self-reported health, number of chronic diseases, an indicator of trust and eligibility for flu vaccination; $e_{it}$ is the error term. We are interested in $\beta_1$ *and* $\beta_2$ capturing the partial correlation of our two measures of computer skills/digital literacy with flu vaccination uptake.

We first provide logit estimates of the equation (1) separately for SCS2 and wave 9. We further test whether the partial correlation of *Pre-pandemic computer skills*$_i$ and *Any pre-pandemic computer job*$_i$ with flu vaccination uptake is statistically different between SCS2 and wave 9 through the following fully-interacted model:

$$y_{it} = \alpha + \delta_1 \text{ Pre-pandemic computer skills}_{it} + \delta_2 \text{ Any pre-pandemic computer job}_{it} + \delta_3 \text{ Pre-pandemic computer skills}_{it}\text{*wave9}_{it} + \delta_4 \text{ Any pre-pandemic computer job}_{it}\text{*wave9}_{it} + \delta_5 \text{ wave9}_{it} + X_1{}'_{it}\eta + e_{it}. \qquad (2)$$

In equation (2) the matrix $X_1{}'_{it}$ contains not only the covariates included in equation (1) but also their interactions with the wave 9 dummy variable. We will estimate also more parsimonious specifications where we (a) do not include the interaction terms between country dummies and the wave 9 binary indicator and (b) include only the interaction terms *Pre-pandemic computer skills*$_{it}$**wave9*$_{it}$ *and Any pre-pandemic computer job*$_{it}$**wave9*$_{it}$.

To account for the fact that we observe the same individuals in SCS2 and wave 9, we cluster standard errors at the individual level.

## 5. Estimates

We report in Table 5 our logit estimates. More precisely, column (1) and column (2) of Table 5 show estimates for SCS2 and wave 9, respectively.

We can see that the partial correlation between our indicators of computer skills/digital literacy is positive, suggesting that individuals having better computer skills are more likely to report being vaccinated against seasonal flu. The estimated marginal effects of *pre-pandemic computer skills* and *any pre-pandemic computer job* are highly significant.

Columns (3) of Table 5 report the logit estimates of the fully-interacted model: we can see that the interaction of *pre-pandemic computer skills* with the dummy wave 9 is positive but not statistically significant at the conventional levels, similarly the interaction term of *pre-pandemic computer job* with the dummy wave 9 is negative but not statistically significant. This result is confirmed also in more parsimonious specifications for the interaction term between *pre-pandemic computer job* and wave 9 (see columns (4) and (5)).

Table 5. Logit estimates of pre-pandemic computer skills/digital literacy on seasonal flu vaccination.

| | (1)<br>SCS2<br>LOGIT<br>(Marginal effects) | (2)<br>Wave 9<br>LOGIT<br>(Marginal effects) | (3)<br>Interacted model<br>LOGIT<br>(Marginal effects) | (4)<br>Interacted model<br>LOGIT 1<br>(Marginal effects) | (5)<br>Interacted model<br>LOGIT 2<br>(Marginal effects) |
|---|---|---|---|---|---|
| Pre-pandemic computer skills | 0.036*** | 0.039*** | 0.037*** | 0.039*** | 0.036*** |
| | (0.008) | (0.008) | (0.008) | (0.008) | (0.008) |
| Pre-pandemic computer skills*wave9 | | | 0.002 | -0.003 | 0.003 |
| | | | (0.008) | (0.008) | (0.007) |
| Any pre-pandemic computer job | 0.031*** | 0.019** | 0.032*** | 0.037*** | 0.033*** |
| | (0.010) | (0.010) | (0.010) | (0.010) | (0.009) |
| Any pre-pandemic computer job*wave9 | | | -0.013 | -0.023** | -0.016** |
| | | | (0.010) | (0.009) | (0.007) |
| Dummy: wave9 | | | 0.445*** | 0.513*** | 0.066*** |
| | | | (0.075) | (0.023) | (0.005) |
| Controls | X | X | X | X | X |
| Country FE | X | X | X | X | X |
| Interactions: Full set of controls* wave9 | | | X | | |
| Interactions: Country dummies* wave9 excluded | | | | X | |
| Interactions: Only pre-pandemic computer skill*wave9 and any pre-pandemic computer job * wave9 | | | | | X |
| Observations | 25, 681 | 25, 681 | 51,362 | 51,362 | 51,362 |
| Log-likelihood | -13931 | -14591 | -28553 | -28663 | -28678 |
| Joint-significance (Chi2) | | | 56.67 | 62.62 | 62.67 |
| Joint-significance (p-value) | | | 0.0000 | 0.0000 | 0.0000 |

Note: In columns (3) to (5) we cluster standard errors at the individual level. Significance level: *** p<0.01, ** p<0.05, * p<0.1

Our results show that individuals with better computer skills/digital literacy are more likely to report being vaccinated against seasonal flu, but we do not estimate differential significant changes in flu vaccination take-up behavior after the pandemic depending on computer skills/digital literacy.

In the Appendix - Table A.1 - we report the full set of estimates, where we can see that age has a non-linear significant effect on seasonal flu vaccination uptake. Being eligible for receiving the vaccination for free is positively correlated with the uptake decision. Better educated individuals as well as individuals with higher household income are more likely to report being vaccinated. Current health, measured through the number of chronic diseases and self-perceived health, is significantly associated with flu vaccination.

## 6. Conclusions

We used data gathered during and after the COVID-19 pandemic through SHARE to provide new evidence on the correlation between computer skills/digital literacy and the seasonal influenza uptake decision and investigate whether there have been behavioural changes for individuals with better computer skills that may have been driven by spillover effects from the pandemic experience.

Our estimates show that the partial correlation between vaccine up-take and our indicators of computer skills/digital literacy is positive, suggesting that individuals having better computer skills are more likely to report being vaccinated against seasonal flu, but we do not estimate significant and sizable differential changes in flu vaccination take-up behavior, after the pandemic, depending on computer skills/digital literacy.

Our results draw the attention on the role of digital inclusion on vaccination take-up behavior among older individuals, suggesting that initiatives that introduce ICT to the elderly population might have positive effects not only to reduce loneliness (Llorente-Barroso et al., 2021), and preventing cognitive decline (Bonilha et al., 2024) but also on vaccination take-up behavior. Promotion of digital inclusion can further improve older people's life quality by allowing them to find health information on the internet and help them better recognizing misinformation.

## References

Binzel, C., & Link, A. (2023, April 24). *DP18109 The Deep Roots of Vaccine Hesitancy in Germany: The 19th-Century Naturopathic Movement*. CEPR. https://cepr.org/publications/dp18109

Blank, P. R., Van Essen, G. A., Ortiz De Lejarazu, R., Kyncl, J., Nitsch-Osuch, A., Kuchar, E. P., Falup-Pecurariu, O., Maltezou, H. C., Zavadska, D., Kristufkova, Z., & Kassianos, G. (2018). Impact of European vaccination policies on seasonal influenza vaccination coverage rates: An update seven years later. *Human Vaccines & Immunotherapeutics*, 1–9. https://doi.org/10.1080/21645515.2018.1489948

Bonilha, A. C., Ribeiro, L. W., Mapurunga, M., Demarzo, M., Ota, F., Andreoni, S., & Ramos, L. R. (2024). Preventing cognitive decline via digital inclusion and virtual game management: An intervention study with older adults in the community. Aging & Mental Health, 28(2), 268–274. https://doi.org/10.1080/13607863.2023.2258104

Börsch-Supan, A., Brandt, M., Hunkler, C., Kneip, T., Korbmacher, J., Malter, F., Schaan, B., Stuck, S., & Zuber, S. (2013). Data Resource Profile: The Survey of Health, Ageing and Retirement in Europe (SHARE). *International Journal of Epidemiology*, *42*(4), 992–1001. https://doi.org/10.1093/ije/dyt088

Davidson, M. (2017). Vaccination as a cause of autism—Myths and controversies. *Dialogues in Clinical Neuroscience*, *19*(4), 403–407. https://doi.org/10.31887/DCNS.2017.19.4/mdavidson

Deiana, C., Geraci, A., Mazzarella, G., & Sabatini, F. (2022). Perceived risk and vaccine hesitancy: Quasi-experimental evidence from Italy. *Health Economics*, *31*(6), 1266–1275. https://doi.org/10.1002/hec.4506

DeStefano, F., & Thompson, W. W. (2004). MMR vaccine and autism: An update of the scientific evidence. *Expert Review of Vaccines*, *3*(1), 19–22. https://doi.org/10.1586/14760584.3.1.19

Dolman, A. J., Fraser, T., Panagopoulos, C., Aldrich, D. P., & Kim, D. (2023). Opposing views: Associations of political polarization, political party affiliation, and social trust with COVID-19 vaccination intent and receipt. *Journal of Public Health*, *45*(1), 36–39. https://doi.org/10.1093/pubmed/fdab401

European Centre for Disease Prevention and Control. (2023). Seasonal influenza vaccination recommendations and coverage rates in EU/EEA Member States: An overview of vaccination recommendations for 2021–22 and coverage rates for the 2018–19 to 2020–21 influenza seasons. Publications Office. https://data.europa.eu/doi/10.2900/335933

Fieselmann, J., Annac, K., Erdsiek, F., Yilmaz-Aslan, Y., & Brzoska, P. (2022). What are the reasons for refusing a COVID-19 vaccine? A qualitative analysis of social media in Germany. *BMC Public Health*, *22*(1), 846. https://doi.org/10.1186/s12889-022-13265-y

Hofstra, L., & Larson, H. J. (2023). Factors Associated With Vaccination Refusal—Critical Lessons From the Omicron Wave in Hong Kong. *JAMA Network Open*, *6*(10), e2337829. https://doi.org/10.1001/jamanetworkopen.2023.37829

Isonne, C., Iera, J., Sciurti, A., Renzi, E., De Blasiis, M. R., Marzuillo, C., Villari, P., & Baccolini, V. (2024). How well does vaccine literacy predict intention to vaccinate and vaccination status? A systematic review and meta-analysis. *Human Vaccines & Immunotherapeutics*, *20*(1), 2300848. https://doi.org/10.1080/21645515.2023.2300848

Isonne, C., Marzuillo, C., Villari, P., & Baccolini, V. (2024). The role of vaccine literacy and health literacy in the health prevention decision-making process. *Human Vaccines & Immunotherapeutics*, *20*(1), 2321675. https://doi.org/10.1080/21645515.2024.2321675

Iuliano, A. D., Roguski, K. M., Chang, H. H., Muscatello, D. J., Palekar, R., Tempia, S., Cohen, C., Gran, J. M., Schanzer, D., Cowling, B. J., Wu, P., Kyncl, J., Ang, L. W., Park, M., Redlberger-Fritz, M., Yu, H., Espenhain, L., Krishnan, A., Emukule, G., … Mustaquim, D. (2018). Estimates of global seasonal influenza-associated respiratory mortality: A modelling study. *The Lancet*, *391*(10127), 1285–1300. https://doi.org/10.1016/S0140-6736(17)33293-2

Jain, L., Vij, J., Satapathy, P., Chakrapani, V., Patro, B., Kar, S. S., Singh, R., Pala, S., Sankhe, L., Modi, B., Bali, S., Rustagi, N., Rajagopal, V., Kiran, T., Goel, K., Aggarwal, A. K., Gupta, M., & Padhi, B. K. (2021). Factors Influencing COVID-19 Vaccination Intentions Among College Students: A Cross-Sectional Study in India. *Frontiers in Public Health*, *9*, 735902. https://doi.org/10.3389/fpubh.2021.735902

Jennings, W., Stoker, G., Bunting, H., Valgarðsson, V. O., Gaskell, J., Devine, D., McKay, L., & Mills, M. C. (2021). Lack of Trust, Conspiracy Beliefs, and Social Media Use Predict COVID-19 Vaccine Hesitancy. *Vaccines*, *9*(6), 593. https://doi.org/10.3390/vaccines9060593

Larson, H. J., Gakidou, E., & Murray, C. J. L. (2022). The Vaccine-Hesitant Moment. *New England Journal of Medicine*, *387*(1), 58–65. https://doi.org/10.1056/NEJMra2106441

Limbu, Y. B., Gautam, R. K., & Pham, L. (2022). The Health Belief Model Applied to COVID-19 Vaccine Hesitancy: A Systematic Review. *Vaccines*, *10*(6), 973. https://doi.org/10.3390/vaccines10060973

Llorente-Barroso, C., Kolotouchkina, O., & Mañas-Viniegra, L. (2021). The Enabling Role of ICT to Mitigate the Negative Effects of Emotional and Social Loneliness of the Elderly during COVID-19 Pandemic. International Journal of Environmental Research and Public Health, 18(8), 3923. https://doi.org/10.3390/ijerph18083923

MacDonald, N. E. & the SAGE Working Group on Vaccine Hesitancy. (2015). Vaccine hesitancy: Definition, scope and determinants. *Vaccine*, *33*(34), 4161–4164. https://doi.org/10.1016/j.vaccine.2015.04.036

Malik, A. A., McFadden, S. M., Elharake, J., & Omer, S. B. (2020). Determinants of COVID-19 vaccine acceptance in the US. *EClinicalMedicine*, *26*, 100495. https://doi.org/10.1016/j.eclinm.2020.100495

Martins Van Jaarsveld, G. (2020). The Effects of COVID-19 Among the Elderly Population: A Case for Closing the Digital Divide. *Frontiers in Psychiatry*, *11*, 577427. https://doi.org/10.3389/fpsyt.2020.577427

McSpadden, J. (2021). *Vaccine Hesitancy among Older Adults, with Implications for COVID-19 Vaccination and Beyond*. AARP Public Policy Institute. https://doi.org/10.26419/ppi.00123.001

Miller, E. A. (Ed.). (2021). *Older Adults and COVID-19* (0 ed.). Routledge. https://doi.org/10.4324/9781003118695

Moon, I., Han, J., & Kim, K. (2023). Determinants of COVID-19 vaccine Hesitancy: 2020 California Health Interview Survey. *Preventive Medicine Reports*, *33*, 102200. https://doi.org/10.1016/j.pmedr.2023.102200

Mustafa, A., Safi, M., Opoku, M. P., & Mohamed, A. M. (2022). The impact of health status on attitudes toward COVID-19 vaccination. *Health Science Reports*, *5*(4), e744. https://doi.org/10.1002/hsr2.744

Nuwarda, R. F., Ramzan, I., Weekes, L., & Kayser, V. (2022). Vaccine Hesitancy: Contemporary Issues and Historical Background. *Vaccines*, *10*(10), 1595. https://doi.org/10.3390/vaccines10101595

Paimre, M., Virkus, S., & Osula, K. (2024). How Technology, Health Information Seeking, and Socioeconomic Factors Are Associated With Coronavirus Disease 2019 Vaccination Readiness in Estonians Over 50 Years? *Health Education & Behavior*, *51*(4), 502–511. https://doi.org/10.1177/10901981241249972

Pertwee, E., Simas, C., & Larson, H. J. (2022). An epidemic of uncertainty: Rumors, conspiracy theories and vaccine hesitancy. *Nature Medicine*, *28*(3), 456–459. https://doi.org/10.1038/s41591-022-01728-z

Principe, F., & Weber, G. (2023). Online health information seeking and Covid-19 vaccine hesitancy: Evidence from 50+ Europeans. *Health Policy*, *138*, 104942. https://doi.org/10.1016/j.healthpol.2023.104942

Richter, L., Schreml, S., & Heidinger, T. (2022a). Ready for Vaccination? COVID-19 Vaccination Willingness of Older People in Austria. *Frontiers in Public Health*, *10*, 859024. https://doi.org/10.3389/fpubh.2022.859024

Richter, L., Schreml, S., & Heidinger, T. (2022b). Ready for Vaccination? COVID-19 Vaccination Willingness of Older People in Austria. *Frontiers in Public Health*, *10*, 859024. https://doi.org/10.3389/fpubh.2022.859024

Rizzo, C., Rezza, G., & Ricciardi, W. (2018). Strategies in recommending influenza vaccination in Europe and US. *Human Vaccines & Immunotherapeutics*, *14*(3), 693–698. https://doi.org/10.1080/21645515.2017.1367463

SHARE-ERIC (2024). Survey of Health, Ageing and Retirement in Europe (SHARE) Wave 1. Release version: 9.0.0. SHARE-ERIC. Data set. DOI: 10.6103/SHARE.w1.900

SHARE-ERIC (2024). Survey of Health, Ageing and Retirement in Europe (SHARE) Wave 2. Release version: 9.0.0. SHARE-ERIC. Data set. DOI: 10.6103/SHARE.w2.900

SHARE-ERIC (2024). Survey of Health, Ageing and Retirement in Europe (SHARE) Wave 3 – SHARELIFE. Release version: 9.0.0. SHARE-ERIC. Data set. DOI: 10.6103/SHARE.w3.900

SHARE-ERIC (2024). Survey of Health, Ageing and Retirement in Europe (SHARE) Wave 4. Release version: 9.0.0. SHARE-ERIC. Data set. DOI: 10.6103/SHARE.w4.900

SHARE-ERIC (2024). Survey of Health, Ageing and Retirement in Europe (SHARE) Wave 5. Release version: 9.0.0. SHARE-ERIC. Data set. DOI: 10.6103/SHARE.w5.900

SHARE-ERIC (2024). Survey of Health, Ageing and Retirement in Europe (SHARE) Wave 6. Release version: 9.0.0. SHARE-ERIC. Data set. DOI: 10.6103/SHARE.w6.900

SHARE-ERIC (2024). Survey of Health, Ageing and Retirement in Europe (SHARE) Wave 7. Release version: 9.0.0. SHARE-ERIC. Data set. DOI: 10.6103/SHARE.w7.900

SHARE-ERIC (2024). Survey of Health, Ageing and Retirement in Europe (SHARE) Wave 8. Release version: 9.0.0. SHARE-ERIC. Data set. DOI: 10.6103/SHARE.w8.900

SHARE-ERIC (2024). Survey of Health, Ageing and Retirement in Europe (SHARE) Wave 9. COVID-19 Survey 2. Release version: 9.0.0. SHARE-ERIC. Data set. DOI: 10.6103/SHARE.w9ca.900

SHARE-ERIC (2024). Survey of Health, Ageing and Retirement in Europe (SHARE) Wave 9. Release version: 9.0.0. SHARE-ERIC. Data set. DOI: 10.6103/SHARE.w9.900

Sharma, N., Basu, S., Lalwani, H., Rao, S., Malik, M., Garg, S., Shrivastava, R., & Singh, M. M. (2023). COVID-19 Booster Dose Coverage and Hesitancy among Older Adults in an Urban Slum and Resettlement Colony in Delhi, India. *Vaccines*, *11*(7), 1177. https://doi.org/10.3390/vaccines11071177

Sherman, S. M., Smith, L. E., Sim, J., Amlôt, R., Cutts, M., Dasch, H., Rubin, G. J., & Sevdalis, N. (2021). COVID-19 vaccination intention in the UK: Results from the COVID-19 vaccination acceptability study (CoVAccS), a nationally representative cross-sectional survey. *Human Vaccines & Immunotherapeutics*, *17*(6), 1612–1621. https://doi.org/10.1080/21645515.2020.1846397

Siram, B., Shah, M., & Panda, R. (2024). Vaccine Hesitancy in COVID-19: A Behavioural Economics Approach—A Systematic Literature Review. *Studies in Microeconomics*, *12*(3), 371–381. https://doi.org/10.1177/23210222221129445

Van Jaarsveld, G. M. (2020). The Effects of COVID-19 Among the Elderly Population: A Case for Closing the Digital Divide. *Frontiers in Psychiatry*, *11*, 577427. https://doi.org/10.3389/fpsyt.2020.577427

Veronese, N., Zambon, N., Noale, M., & Maggi, S. (2023). Poverty and Influenza/Pneumococcus Vaccinations in Older People: Data from The Survey of Health, Ageing and Retirement in Europe (SHARE) Study. *Vaccines*, *11*(9), 1422. https://doi.org/10.3390/vaccines11091422

Wolff, K. (2021). COVID-19 Vaccination Intentions: The Theory of Planned Behavior, Optimistic Bias, and Anticipated Regret. *Frontiers in Psychology*, *12*, 648289. https://doi.org/10.3389/fpsyg.2021.648289

World Health Organization (2019). Ten threats to global health in 2019. World Health Organization. https://www.who.int/news-room/spotlight/ten-threats-to-global-health-in-2019

World Health Organization (2023). Influenza (seasonal). World Health Organization. https://www.who.int/news-room/fact-sheets/detail/influenza-(seasonal)

World Health Organization (2022). Vaccine against influenza; WHO position paper – May 2022. Weekly epidemiological record. 97(19), 185-208. https://www.who.int/publications/i/item/who-wer9719

World Health Organization (2022). Understanding the behavioural and social drivers of vaccine uptake; WHO position paper – May 2022. Weekly epidemiological record. 97(20), 209-224. https://www.who.int/publications/i/item/who-wer9720

Xie, B., Charness, N., Fingerman, K., Kaye, J., Kim, M. T., & Khurshid, A. (2020). When Going Digital Becomes a Necessity: Ensuring Older Adults' Needs for Information, Services, and Social Inclusion During COVID-19. Journal of Aging & Social Policy, 32(4–5), 460–470. https://doi.org/10.1080/08959420.2020.1771237

Yasmin, F., Najeeb, H., Moeed, A., Naeem, U., Asghar, M. S., Chughtai, N. U., Yousaf, Z., Seboka, B. T., Ullah, I., Lin, C.-Y., & Pakpour, A. H. (2021). COVID-19 Vaccine Hesitancy in the United States: A Systematic Review. *Frontiers in Public Health*, *9*, 770985. https://doi.org/10.3389/fpubh.2021.770985

Zapletal, A., Wells, T., Russell, E., & Skinner, M. W. (2023). On the triple exclusion of older adults during COVID-19: Technology, digital literacy and social isolation. *Social Sciences & Humanities Open*, *8*(1), 100511. https://doi.org/10.1016/j.ssaho.2023.100511

Zhang, Y., Zhang, L., Guan, H., Hao, R., & Liu, W. (2024). The impact of internet health information seeking on COVID-19 vaccination behavior in China. *BMC Public Health*, *24*(1), 89. https://doi.org/10.1186/s12889-024-17638-3

Zheng, H., Jiang, S., & Rosenthal, S. (2022). Linking Online Vaccine Information Seeking to Vaccination Intention in the Context of the COVID-19 Pandemic. *Science Communication*, *44*(3), 320–346. https://doi.org/10.1177/10755470221101067

Zintel, S., Flock, C., Arbogast, A. L., Forster, A., Von Wagner, C., & Sieverding, M. (2023). Gender differences in the intention to get vaccinated against COVID-19: A systematic review and meta-analysis. *Journal of Public Health*, *31*(8), 1303–1327. https://doi.org/10.1007/s10389-021-01677-w

# Appendix

## Table A1. Logit estimates. Dep.Var. Seasonal flu vaccination take-up.

| | (1) SCS2 LOGIT (marginal effects) | (2) Wave 9 LOGIT (marginal effects) | (3) Interacted model LOGIT (marginal effects) | (4) Interacted model LOGIT 1 (marginal effects) | (5) Interacted model LOGIT 2 (marginal effects) |
|---|---|---|---|---|---|
| **Dummy: wave9** | | | 0.445*** | 0.513*** | 0.066*** |
| | | | (0.075) | (0.023) | (0.005) |
| **Pre-pandemic computer skill: Good** | 0.036*** | 0.039*** | 0.037*** | 0.039*** | 0.036*** |
| | (0.008) | (0.008) | (0.008) | (0.008) | (0.008) |
| 1.comp_skill_prep#1.wave9 | | | 0.002 | -0.003 | 0.003 |
| | | | (0.008) | (0.008) | (0.007) |
| **Any pre-pandemic computer job: Yes** | 0.031*** | 0.019** | 0.032*** | 0.037*** | 0.033*** |
| | (0.010) | (0.010) | (0.010) | (0.010) | (0.009) |
| 1.p_comp_job2#1.wave9 | | | -0.013 | -0.023** | -0.016** |
| | | | (0.010) | (0.009) | (0.007) |
| **Eligible for flu vaccination in the current wave: Yes** | 0.079*** | 0.085*** | 0.081*** | 0.072*** | 0.083*** |
| | (0.012) | (0.012) | (0.013) | (0.013) | (0.011) |
| 1.eligibility_current#1.wave9 | | | 0.001 | 0.021* | |
| | | | (0.012) | (0.011) | |
| ***Health variables*** | | | | | |
| **Numbers of chronic diseases** | 0.042*** | 0.047*** | 0.043*** | 0.041*** | 0.043*** |
| | (0.003) | (0.003) | (0.003) | (0.003) | (0.002) |
| 1.wave9#c.chronics | | | 0.002 | 0.004 | |
| | | | (0.003) | (0.003) | |
| **Is trustful toward others** | 0.004*** | 0.005*** | 0.004*** | 0.002 | 0.004*** |
| | (0.001) | (0.001) | (0.001) | (0.001) | (0.001) |
| 1.wave9#c.trust | | | 0.001 | 0.005*** | |
| | | | (0.001) | (0.001) | |
| **Subjective health: Poor** | -0.033*** | -0.033*** | -0.034*** | -0.036*** | -0.029*** |
| | (0.013) | (0.010) | (0.013) | (0.013) | (0.007) |
| 1.sr_health#1.wave9 | | | 0.002 | 0.011 | |
| | | | (0.016) | (0.017) | |
| ***Socio-economic variables*** | | | | | |
| **Is a female: Yes** | 0.009* | 0.007 | 0.010* | 0.010* | 0.008* |
| | (0.006) | (0.006) | (0.006) | (0.006) | (0.005) |
| 1.female#1.wave9 | | | -0.002 | -0.003 | |
| | | | (0.005) | (0.006) | |
| **Age** | 0.012*** | 0.002 | 0.012*** | 0.015*** | 0.006** |
| | (0.004) | (0.003) | (0.004) | (0.004) | (0.003) |
| 1.wave9#c.age | | | -0.010*** | -0.015*** | |
| | | | (0.004) | (0.004) | |
| **Age squared** | 0.015 | 0.052*** | 0.015 | 0.005 | 0.036*** |
| | (0.017) | (0.014) | (0.017) | (0.018) | (0.013) |
| 1.wave9#c.age_sq | | | 0.036** | 0.053*** | |
| | | | (0.015) | (0.016) | |
| **Age cubed** | -0.006*** | -0.010*** | -0.006*** | -0.005** | -0.008*** |
| | (0.002) | (0.002) | (0.002) | (0.002) | (0.002) |
| 1.wave9#c.age_cb | | | -0.004* | -0.006** | |
| | | | (0.002) | (0.002) | |
| **Highest education attained (ISCED)** *Ref: None* | | | | | |
| Primary education | 0.037** | 0.023 | 0.038** | 0.036* | 0.030* |
| | (0.019) | (0.019) | (0.019) | (0.019) | (0.016) |
| Secondary education | 0.025 | 0.027 | 0.026 | 0.023 | 0.026 |
| | (0.019) | (0.019) | (0.019) | (0.019) | (0.016) |
| Tertiary education | 0.054*** | 0.065*** | 0.055*** | 0.051** | 0.060*** |
| | (0.020) | (0.021) | (0.020) | (0.020) | (0.017) |
| 2.yedu_isced11#1.wave9 | | | -0.015 | -0.011 | |
| | | | (0.020) | (0.020) | |
| 3.yedu_isced11#1.wave9 | | | -0.000 | 0.006 | |
| | | | (0.020) | (0.020) | |
| 4.yedu_isced11#1.wave9 | | | 0.008 | 0.017 | |
| | | | (0.021) | (0.021) | |
| **Area of house building: Urban** | 0.013** | 0.021*** | 0.014** | 0.011* | 0.017*** |
| | (0.007) | (0.007) | (0.007) | (0.007) | (0.006) |
| 1.urban#1.wave9 | | | 0.007 | 0.011* | |
| | | | (0.006) | (0.006) | |
| **Household size** | -0.005 | -0.005 | -0.005 | -0.004 | -0.005 |
| | (0.003) | (0.004) | (0.004) | (0.003) | (0.003) |
| 1.wave9#c.hhsize | | | 0.000 | -0.002 | |
| | | | (0.004) | (0.004) | |
| **Household's income quartile during wave 8** | | | | | |
| *Ref: First quartile* | | | | | |
| Second quartile | 0.053*** | 0.048*** | 0.054*** | 0.054*** | 0.051*** |
| | (0.008) | (0.008) | (0.008) | (0.008) | (0.007) |
| Third quartile | 0.055*** | 0.052*** | 0.056*** | 0.057*** | 0.054*** |
| | (0.008) | (0.009) | (0.009) | (0.008) | (0.007) |
| Fourth quartile | 0.070*** | 0.059*** | 0.071*** | 0.073*** | 0.064*** |
| | (0.009) | (0.009) | (0.009) | (0.009) | (0.008) |
| 2.thinc2q_w8#1.wave9 | | | -0.007 | -0.006 | |
| | | | (0.008) | (0.008) | |
| 3.thinc2q_w8#1.wave9 | | | -0.006 | -0.007 | |
| | | | (0.008) | (0.008) | |
| 4.thinc2q_w8#1.wave9 | | | -0.014 | -0.016* | |
| | | | (0.009) | (0.009) | |
| **Never did any paid work before the pandemic: Yes** | 0.009 | 0.009 | 0.009 | 0.011 | 0.009 |
| | (0.009) | (0.009) | (0.009) | (0.009) | (0.008) |
| 1.pp_neverpaidwork#1.wave9 | | | -0.000 | -0.003 | |
| | | | (0.009) | (0.009) | |
| **Country of residence** | | | | | |
| *Ref: Austria* | | | | | |
| 12, Germany | 0.228*** | 0.234*** | 0.230*** | 0.234*** | 0.233*** |
| | (0.016) | (0.015) | (0.016) | (0.014) | (0.014) |
| 13, Sweden | 0.210*** | 0.305*** | 0.212*** | 0.260*** | 0.260*** |
| | (0.019) | (0.017) | (0.019) | (0.016) | (0.016) |
| 14, Netherlands | 0.289*** | 0.313*** | 0.289*** | 0.305*** | 0.304*** |
| | (0.024) | (0.023) | (0.024) | (0.021) | (0.021) |
| 15, Spain | 0.281*** | 0.310*** | 0.282*** | 0.299*** | 0.299*** |
| | (0.020) | (0.018) | (0.020) | (0.017) | (0.017) |
| 16, Italy | 0.229*** | 0.298*** | 0.231*** | 0.266*** | 0.266*** |
| | (0.017) | (0.015) | (0.017) | (0.014) | (0.014) |
| 17, France | 0.246*** | 0.225*** | 0.247*** | 0.238*** | 0.238*** |
| | (0.016) | (0.016) | (0.016) | (0.014) | (0.014) |
| 18, Denmark | 0.287*** | 0.367*** | 0.288*** | 0.331*** | 0.330*** |
| | (0.017) | (0.014) | (0.016) | (0.013) | (0.013) |
| 19, Greece | 0.232*** | 0.306*** | 0.236*** | 0.272*** | 0.272*** |
| | (0.016) | (0.015) | (0.016) | (0.013) | (0.013) |
| 20, Switzerland | 0.067*** | 0.063*** | 0.068*** | 0.065*** | 0.065*** |
| | (0.017) | (0.017) | (0.017) | (0.015) | (0.015) |
| 23, Belgium | 0.283*** | 0.270*** | 0.283*** | 0.279*** | 0.279*** |
| | (0.016) | (0.015) | (0.016) | (0.014) | (0.014) |
| 25, Israel | 0.212*** | 0.295*** | 0.214*** | 0.256*** | 0.256*** |
| | (0.027) | (0.025) | (0.027) | (0.023) | (0.023) |
| 28, Czech Republic | -0.045*** | -0.052*** | -0.047*** | -0.049*** | -0.049*** |
| | (0.017) | (0.017) | (0.017) | (0.015) | (0.015) |
| 29, Poland | -0.209*** | -0.176*** | -0.219*** | -0.193*** | -0.193*** |
| | (0.016) | (0.018) | (0.018) | (0.015) | (0.015) |
| 31, Luxembourg | 0.189*** | 0.189*** | 0.191*** | 0.191*** | 0.191*** |
| | (0.021) | (0.020) | (0.021) | (0.019) | (0.019) |
| 32, Hungary | 0.119 | -0.078 | 0.121 | 0.029 | 0.028 |
| | (0.140) | (0.179) | (0.141) | (0.121) | (0.120) |
| 34, Slovenia | -0.014 | -0.047*** | -0.014 | -0.030** | -0.030** |
| | (0.017) | (0.017) | (0.017) | (0.015) | (0.015) |
| 35, Estonia | -0.159*** | -0.003 | -0.164*** | -0.076*** | -0.075*** |
| | (0.015) | (0.016) | (0.015) | (0.014) | (0.014) |
| 47, Croatia | 0.056*** | 0.049** | 0.057*** | 0.053*** | 0.053*** |
| | (0.020) | (0.020) | (0.020) | (0.017) | (0.017) |
| 48, Lithuania | -0.097*** | -0.118*** | -0.100*** | -0.108*** | -0.108*** |
| | (0.019) | (0.020) | (0.020) | (0.017) | (0.017) |
| 51, Bulgaria | -0.309*** | -0.145*** | -0.332*** | -0.221*** | -0.221*** |
| | (0.017) | (0.025) | (0.020) | (0.018) | (0.018) |
| 53, Cyprus | 0.224*** | 0.204*** | 0.226*** | 0.217*** | 0.217*** |
| | (0.029) | (0.028) | (0.028) | (0.026) | (0.026) |
| 55, Finland | 0.244*** | 0.261*** | 0.245*** | 0.255*** | 0.255*** |
| | (0.019) | (0.018) | (0.019) | (0.017) | (0.017) |
| 57, Latvia | -0.161*** | -0.135*** | -0.168*** | -0.149*** | -0.149*** |
| | (0.022) | (0.024) | (0.024) | (0.019) | (0.019) |
| 59, Malta | 0.358*** | 0.308*** | 0.355*** | 0.337*** | 0.336*** |
| | (0.019) | (0.019) | (0.018) | (0.017) | (0.017) |
| 61, Romania | 0.027 | 0.052** | 0.028 | 0.039** | 0.039** |
| | (0.021) | (0.021) | (0.021) | (0.018) | (0.018) |
| 63, Slovakia | -0.216*** | -0.036 | -0.227*** | -0.114*** | -0.114*** |
| | (0.021) | (0.023) | (0.023) | (0.018) | (0.018) |
| 12.country#1.wave9 | | | 0.002 | | |
| | | | (0.015) | | |
| 13.country#1.wave9 | | | 0.102*** | | |
| | | | (0.021) | | |
| 14.country#1.wave9 | | | 0.029 | | |
| | | | (0.024) | | |
| 15.country#1.wave9 | | | 0.033 | | |
| | | | (0.024) | | |
| 16.country#1.wave9 | | | 0.072*** | | |
| | | | (0.019) | | |
| 17.country#1.wave9 | | | -0.024 | | |
| | | | (0.015) | | |
| 18.country#1.wave9 | | | 0.099*** | | |
| | | | (0.019) | | |
| 19.country#1.wave9 | | | 0.076*** | | |
| | | | (0.018) | | |
| 20.country#1.wave9 | | | -0.006 | | |
| | | | (0.016) | | |
| 23.country#1.wave9 | | | -0.016 | | |
| | | | (0.015) | | |
| 25.country#1.wave9 | | | 0.090*** | | |
| | | | (0.030) | | |
| 28.country#1.wave9 | | | -0.004 | | |
| | | | (0.016) | | |
| 29.country#1.wave9 | | | 0.063*** | | |
| | | | (0.023) | | |
| 31.country#1.wave9 | | | -0.003 | | |
| | | | (0.019) | | |
| 32.country#1.wave9 | | | -0.180 | | |
| | | | (0.159) | | |
| 34.country#1.wave9 | | | -0.031** | | |
| | | | (0.016) | | |
| 35.country#1.wave9 | | | 0.168*** | | |
| | | | (0.017) | | |
| 47.country#1.wave9 | | | -0.009 | | |
| | | | (0.019) | | |
| 48.country#1.wave9 | | | -0.015 | | |
| | | | (0.020) | | |
| 51.country#1.wave9 | | | 0.287*** | | |
| | | | (0.044) | | |
| 53.country#1.wave9 | | | -0.024 | | |
| | | | (0.025) | | |
| 55.country#1.wave9 | | | 0.016 | | |
| | | | (0.019) | | |

| | (1) | (2) | (3) | (4) | (5) |
|---|---|---|---|---|---|
| 57.country#1.wave9 | | | 0.045 | | |
| | | | (0.031) | | |
| 59.country#1.wave9 | | | -0.057*** | | |
| | | | (0.020) | | |
| 61.country#1.wave9 | | | 0.023 | | |
| | | | (0.021) | | |
| 63.country#1.wave9 | | | 0.218*** | | |
| | | | (0.031) | | |
| Observations | 25, 681 | 25, 681 | 51,362 | 51,362 | 51,362 |
| Log-likelihood | -13931.10 | -14591.52 | -28552.62 | -28662.78 | -28677.78 |
| Joint-significance (Chi2) | | | 56.67 | 62.62 | 62.67 |
| Joint-significance (p-value) | | | 0.0000 | 0.0000 | 0.0000 |

Note: In columns (3) to (5) we cluster standard errors at the individual level. Significance level: *** p<0.01, ** p<0.05, * p<0.1